# Low-loss, silicon integrated, aluminum nitride photonic circuits and their use for electro-optic signal processing

Chi Xiong[1], Wolfram H. P. Pernice[1,†], and Hong X. Tang[1,*]

[1]Department of Electrical Engineering, Yale University, New Haven, CT 06511, USA
† Current address: Institute of Nanotechnology, Karlsruhe Institute of Technology, 76344 Eggenstein-Leopoldshafen, Germany
* Corresponding author electronic mail: hong.tang@yale.edu

**Abstract:** Photonic miniaturization requires seamless integration of linear and nonlinear optical components to achieve passive and active functions simultaneously. Among the available material systems, silicon photonics holds immense promise for optical signal processing and on-chip optical networks. However, silicon is limited to wavelengths above 1.1 μm and does not provide the desired lowest order optical nonlinearity for active signal processing. Here we report the integration of aluminum nitride (AlN) films on silicon substrates to bring active functionalities to chip-scale photonics. Using CMOS-compatible sputtered thin films we fabricate AlN-on-insulator waveguides that exhibit low propagation loss (0.6 dB/cm). Exploiting AlN's inherent Pockels effect we demonstrate electro-optic modulation up to 4.5 Gb/s with very low energy consumption (down to 10 fJ/bit). The ultra-wide transparency window of AlN devices also enables high speed modulation at visible wavelengths. Our low cost, wideband, carrier-free photonic circuits hold promise for ultra-low power and high speed signal processing at the microprocessor chip level.

Photonic integrated circuits offer excellent prospects for high speed complex optical applications on a chip scale. During the past decade, silicon-on-insulator (SOI) has emerged as one of the most promising integrated optics platforms due to the high refractive index contrast it provides, which allows for building ultra-compact devices[1-6]. However, silicon has a narrow indirect bandgap (1.1 eV) and centrosymmetric crystal structure, which not only limits its operation to wavelengths above 1100 nm but also precludes important active functionalities, such as light emission, second harmonic generation and the linear electro-optic (Pockels) effect. $\chi^{(2)}$ nonlinearity is particularly important for wavelength conversion and also the basis of the electro-optic effect, which is commonly employed for high-speed modulation. In order to implement optical emitters, significant progress has been recently made through heterogeneous integration of III-V epitaxial films on silicon[7], in spite of inherent challenges in material sciences and device engineering. To achieve modulation in silicon devices, the plasma dispersion effect is most commonly exploited to manipulate the refractive index of silicon waveguides through mechanisms such as carrier accumulation, carrier injection and depletion[8-11]. Due to the extended lifetime of the free carriers, a careful balance between the modulation speed, efficiency, doping configuration and carrier induced absorption has to be maintained. The Pockels effect, on the other hand, is a wide-band, intrinsic property of non-centrosymmetric crystals. Modulation based on the Pockels effect does not rely on free carriers and has been widely employed in commercial high speed electro-optic modulators, such as devices made from $LiNbO_3$[12] and electro-optic organics[13, 14]. However, common electro-optic materials with large Pockels coefficient are usually not compatible with CMOS semiconductor technologies. Within this context, the search for a suitable electro-optic material with potential of large scale integration on silicon substrates remains an open issue.

Here, we present a nanophotonic circuit system based on sputter-deposited aluminum nitride (AlN) films to meet the demand for high quality Pockels materials on silicon substrates. AlN has found a wide range of applications in ultra-violet (UV) optoelectronics[15] and micro-electro-mechanical-systems (MEMS)[16] because of its direct bandgap and high piezoelectric coupling efficiency. For integrated optics, we introduce AlN as an attractive electro-optic material because of its significant second order nonlinearity and mature deposition technology compatible with a variety of substrates including silicon. Moreover, AlN has one of the largest bandgap (6.2 eV) among all known semiconductors, and thus it not only provides suppression of two photon absorption but also allows for wide band operation from UV to infrared wavelengths. Additionally, compared with silicon, AlN has a superior thermal conductivity ($\kappa_{AlN}$=285 W/m·K) and a small thermo-optic coefficient[17] ($dn_{AlN}/dT$=2.32×10$^{-5}$/K), which is why AlN has been used as electronic substrates and chip carriers where high thermal conductivity is essential. Hence, AlN photonic devices are expected to be more tolerant to temperature fluctuations as well.

In this study, we develop AlN-on-insulator substrates using sputter-deposited AlN films as the light guiding medium. Waveguides fabricated on this platform display propagation loss as low as 0.6 dB/cm at telecom wavelengths. To demonstrate the intrinsic linear electro-optic (Pockels) effect, we fabricate and measure electrically tunable microring resonators for both telecom wavelengths (1550 nm) and visible wavelengths (770 nm). The prototype telecom band microring modulators show none-return-to-zero (NRZ) modulation up to 4.5 Gb/s, currently limited by the cavity photon lifetime.

Wafer scale AlN deposition technology has been widely studied due to demand for applications in bulk acoustic devices (BAW) and piezoelectric actuators[18]. Among the available techniques, radio frequency magnetron reactive sputtering has been established as a leading technology to achieve high quality AlN films with excellent surface morphology[19, 20]. Sputter-deposited AlN thin films are also excellent candidates for integrated optics for a number of reasons. First, with optimized processes the sputter-deposited AlN films feature a columnar micrograin structure with c-axis orientation ([0002] direction) normal to the film plane as shown in Fig.1(a). This preferred c-axis orientation is essential for exploitation of the largest component of AlN's $\chi^{(2)}$ tensor ($d_{33}$). The quality of AlN's c-axis orientation is characterized by the X-ray diffraction rocking curve measurement for the AlN (0002) peak. As shown in inset of Fig 1(b), the typical full width at half maximum (FWHM) for our AlN films is less than 2°, indicating that the film is highly oriented. Second, unlike silicon nitride, AlN can be deposited with residual stress as low as ±75 MPa, allowing for micrometer thick films to be realized and thus permitting the fabrication of high confinement waveguides in both transverse-electric (TE) and transverse-magnetic(TM) polarizations. Third, during the sputtering process, the equilibrium wafer temperature will not exceed a relatively low temperature (350°C) compared with furnace based chemical vapor deposition (normally >1000°C). As a result, the sputtering process of AlN is thermally compatible with critical CMOS processes in which no high temperature can be tolerated. And last, AlN can be sputter-deposited on a variety of crystalline and amorphous CMOS materials, such as silicon, silicon dioxide, silicon nitride and metals. Excellent surface smoothness can therefore be achieved provided that starting substrates have smooth surface morphology.

Here we utilize AlN thin films (ordinary refractive index $n_o$=2.12 at 1550 nm) on top of a low refractive index silicon dioxide (n=1.46) buffer layer, thermally grown on bare 100 mm silicon wafers. AlN films are deposited in a dual cathode ac (40 kHz) powered S-gun magnetron sputtering system. Pure aluminum (99.999%) targets are sputtered in an argon and nitrogen gas mixture. The flow rates of nitrogen and argon gases are optimized to be around 20 sccm and 4 sccm, respectively, to yield best stoichiometry and crystal orientation. The AlN films are deposited without external heating. By adjusting the RF bias power, the stress in the films can be tailored to be around ± 75 MPa. Nanophotonic waveguides are patterned using electron beam lithography on a Vistec EBPG 5000+ 100kV writer. Following lithography in hydrogen silsesquioxane (HSQ) resist, the exposed structures are transferred into the AlN thin films using inductively coupled plasma (ICP) reactive ion etching in $Cl_2$/$BCl_3$/Ar chemistry (Oxford PlasmaLab100). $Cl_2$, $BCl_3$ and Ar flow at a rate of 25 sccm, 9 sccm and 6 sccm, respectively. The measured etch rate is about 200 nm/min at a RF bias power of 70 W and ICP power of 360 W under 5 mTorr pressure. The etching selectivity of the AlN film to the HSQ resist is about 2:1. In Fig 1 (b) we show a cross-sectional scanning electron microscope (SEM) image of a cleaved section of the photonic circuit to illustrate the AlN waveguide geometry sitting on top of the thermal oxide layer. We measure low root-mean-square roughness of 1.5 nm on the film's top surface resulting from the smoothness of the starting $SiO_2$ surface. The deposited AlN film is 650 nm thick and a typical width of the waveguide is 1 μm, permitting telecom band light to be highly confined. The thickness of the thermally grown $SiO_2$ is chosen to be 2 μm, optimized for the highest out-of-plane fiber-to-chip coupling efficiency with focusing grating couplers.

Nonlinear optical applications often involve long propagation distances or alternatively rely on high Q optical resonators for enhancement of light-matter interactions. As such, low loss propagation becomes of fundamental importance to meet the demand for longer effective waveguide length. To characterize the loss performance of our AlN waveguides, we fabricate identical microring resonators (as shown in Fig. 1(c)) with a diameter of 80 μm and different coupling gaps from 200 nm to 600 nm. By varying the gap between the feeding waveguide and the microring resonator, the coupling conditions can be tuned to achieve both critical coupling with high extinction ratio and weak coupling for wider separation. Light from a tunable laser source (New Focus 6428) is coupled into and out of the waveguides using an array of cleaved single-mode polarization-maintaining fibers aligned to the input/output grating couplers. The typical coupling efficiency is 30% per grating coupler for 1550 nm light. The transmission past the microring resonator is recorded with a low-noise photo-receiver (New Focus 2011). As shown in Fig 1 (d), with TE-polarized light input, the transmission spectrum of the device with a coupling gap of 300 nm shows optical resonances separated by a free spectral range of 4.3 nm, corresponding to a group index of 2.01. For devices with an extinction ratio of 15 dB we find a loaded quality factor of 100,000 from a Lorentzian fit to the resonance, signifying that the resonator is operated in the slightly overcoupled regime. When the coupling gap is increased to 800 nm, the ring is operated in the under-coupled regime and the quality factor improves to 600,000, which is expected to be closer to its intrinsic value. Using the expression $\alpha = 10\log_{10} e \cdot 2\pi n_g / (Q_{int} \cdot \lambda_0)$, ($Q_{int}$ is the intrinsic quality factor, $n_g$ is the group index, $\lambda_0$ is the wavelength), we determine a propagation loss of 0.6 dB/cm, which is on par with the performance of state-of-the-art SOI waveguides[21, 22].

In addition to excellent linear optical performance, AlN photonic circuits offer electro-optic effects which can be used to enhance passive optical devices with active functions. Indeed, optical modulation is one of the main required functionalities for the implementation of chip-level optical interconnects[23]. Integrated optical modulators will ideally have a compact footprint and be able to operate at high speed with low power consumption. Because the linear electro-optic (Pockels) effect is intrinsically wideband, integration of electro-optic crystals on silicon substrates is a very attractive option for next generation silicon photonics. Additionally, DC leakage power is entirely eliminated due to the insulating nature of wide bandgap crystals. Furthermore, compared with silicon modulators, crystal-based modulators can have much lower device complexity by eliminating layers of masks for different doping and vias and hence reduce the cost of fabrication.

The relevant electro-optic coefficient of AlN ($r_{33}$, $r_{13}$ ~1 pm/V) [24], is comparable with that of GaAs ($r_{14}$=1.5 pm/V) [25] which has been widely employed in commercial phase and amplitude modulators[26, 27]. Compared with LiNbO$_3$ which has a higher electro-optic coefficient ($r_{33}$=33 pm/V), low cost, the potential for chip-scale integration and CMOS compatibility of AlN compensate for its lower EO coefficient. Further, by employing nanophotonic waveguides, the electro-optic interaction is expected to be enhanced by highly confined optical and electric fields.

Single crystal AlN is a uniaxial material with 6mm symmetry and c-axis as the optic axis. Therefore the electro-optic coefficient matrix of AlN only provides non-vanishing elements of *r$_{13}$, r$_{33}$* and *r$_{51}$*. The refractive index change due to the presence of an applied electrical field is then given by the following equation

$$\Delta\left(\frac{1}{n^2}\right)_i = \begin{pmatrix} 0 & 0 & r_{13} \\ 0 & 0 & r_{13} \\ 0 & 0 & r_{33} \\ 0 & r_{51} & 0 \\ r_{51} & 0 & 0 \\ 0 & 0 & 0 \end{pmatrix} \begin{pmatrix} E_x \\ E_y \\ E_z \end{pmatrix} \quad (1)$$

In the above equation, $E_x$, $E_y$ are the in-plane components of the applied RF field, and $E_z$ is the out-of-plane component of the applied RF field (the coordinate system is illustrated in Fig 2(a) inset). Using the above expression, the refractive indices for the propagating modes in both TE polarization ($n_{x,y}$) and TM polarization ($n_z$) can be obtained. Given that the applied RF electric field is oriented parallel to the c-axis ($E_z$), the resulting refractive indices are given as:

$$n_{x,y} = n_o - (1/2) r_{13} n_o^3 E_z$$
$$n_z = n_e - (1/2) r_{33} n_e^3 E_z \quad (2)$$

where $n_o$ and $n_e$ are the ordinary and extraordinary refractive indices, respectively. When on the other hand the electric field is oriented perpendicular to the c-axis ($E_{x,y}$), the index ellipsoid will have rotated principal axes in $x´$, $y´$ and $z´$. The refractive indices for TE-polarized light and TM-polarized light are then given as:

$$n_{x',y'} = n' + (1/2) n'^3 r_{51} E_{x,y}$$
$$n_{z'} = n' - (1/2) n'^3 r_{51} E_{x,y} \quad (3)$$

where $n' = n_o + 1/2(n_e - n_o)$.

It is notable that in both cases, the RF field dependent refractive index for both polarizations is identical in sign as well as magnitude regardless of the in-plane direction. In other words, the crystallographic orientation in the *x-y* direction does not alter the effective electro-optic tensor formulation. Because our polycrystalline AlN films have c-axis out-of-plane orientation while maintaining in-plane isotropy, the electro-optic matrix in Eq. (1) remains valid to describe the effective electro-optic response of our polycrystalline films.

Equation (1) also indicates that, in order to access AlN's largest EO coefficient[24] $r_{13}$ and $r_{33}$, an out-of-plane electric field ($E_z$) is required, which is normally harder to obtain with planar electrodes. To introduce an effective out-of-plane electric field ($E_z$) overlapping with the guided optical mode, we deposit a layer of $SiO_2$ using plasma-enhanced chemical vapor deposition (PECVD) as a cladding layer on top of the waveguides. A set of ground-signal-ground (GSG) contact pads are placed atop the PECVD oxide. Fig 2 (a) and (b) illustrate the numerically calculated fundamental TE optical mode profile and the electric field distribution respectively. From the finite element simulations we find with the top electrode configuration a significant $E_z$ component across the AlN waveguide cross section. Since the $E_z$ component of the electric field contributes most effectively to the modulation, it is desirable to fabricate the center electrode close to the AlN waveguides so that a larger electro-optic overlap integral $\Gamma$ is achieved. (The overlap integral is defined as $\Gamma = \frac{g}{V} \frac{\iint E_{x,op}^2 E_z dxdz}{\iint E_{x,op}^2 dxdz}$, where $g$ is the electrode gap, $V$ is the voltage, $E_{x,op}$ is the optical field, $E_z$ is the RF field). As shown in the simulation results (Fig 2(c)), a trade-off exists between the maximization of the electro-optic overlap integral (red line) and avoidance of excessive optical absorption from the center electrode (orange line). The simulation also suggests that the average electric field ($E_{avg} = \frac{\iint E_{x,op}^2 E_z dxdz}{\iint E_{x,op}^2 dxdz}$) inside the AlN waveguides increases as the electrode gap becomes smaller (Fig 2(d)). In our experiment, we choose a deposited oxide thickness of 0.8 µm and electrode gap of 2 µm to allow for minimum absorption loss (<1 dB/cm) and a comfortable lateral alignment tolerance.

After the $SiO_2$ deposition, the GSG metallic Ti/Au electrodes (10 nm/200 nm) were deposited on top of the $SiO_2$ cladding layer and patterned by a PMMA ebeam lithography recipe and subsequent lift-off procedure. Fig 2 (e) shows an optical micrograph of a fabricated AlN microring modulator circuit with optical input/output ports in the bottom half and RF electrodes in the top half. When a DC bias voltage is applied on the electrodes, the resonance conditions for the microring resonator will change in response to the modified refractive index. As shown in Fig 2 (f), we measure a near critically coupled microring TE-mode resonance around 1542.10 nm with extinction ratio of 10 dB. The resonance shifts 8 pm when the bias voltage is varied from -7.5 V to +7.5 V.

Additionally, to confirm the assumption that the out-of-plane electric field allows for the largest EO coefficients $r_{13}$, $r_{33}$ to be used, we fabricate microring resonators with two in-plane electrodes (as illustrated in the inset of Fig 2 (g).) With a pair of ground and signal electrodes placed next to the sides of the AlN waveguide, we are able to introduce a lateral electric field ($E_x$) and hence probe the $r_{51}$ EO coefficient. We measure the response of a microring resonator around one of its resonance at 1556.5 nm. As shown in Fig 2 (g), with ±7.5 V voltage bias, the resonance only shifts by about 1 pm, which is close to the measurement error level. Considering the effective magnitude of electric field inside the waveguides, we estimate that $r_{51}$ is at least one order of magnitude smaller than $r_{13}$.

To characterize the micro-ring modulator's capability for carrying out digital transmission at high frequency, the input light's wavelength is tuned into one of the ring resonator's resonances near 1540 nm with a Q of 80,000. For maximum modulation transduction, the wavelength is biased at half of the transmission point near the resonance, corresponding to the location of the largest slope. A 32-bit binary non-return-to-zero (NRZ) test sequence at 1 Gb/s with a peak to peak voltage ($V_{pp}$) of 6.6V (Fig. 3 (a)) is applied on the electrodes. The output light from the grating coupler is collected with a 1 GHz InGaAs photoreceiver. Fig 3 (b) shows the normalized output waveform as recorded by a digital oscilloscope, showing that the modulator correctly modulates the light intensity according to the applied digital sequence.

For optical communication purposes, ultra-fast modulators operating at multi-gigabit per second data rates are required. To test our AlN micro-ring modulator for high data rate operation, a $2^{23}$-1 NRZ pseudo-random binary sequence (PRBS) from a high speed pattern generator (Anritsu MP1601) is applied to the modulator, and the output light from the device is sent to a erbium doped fiber optical amplifier (Pritel FA-20) and a 6 GHz photodetector (HP83440B). With a modulating peak-to-peak voltage V$_{pp}$ of 4V, the eye diagram exhibits clear eye openings at 2.5 Gb/s with an extinction ratio of 3 dB, as shown in Fig. 3 (d). The eye opening and extinction ratio do not exhibit apparent degradation up to 4.5 Gb/s, as shown in Fig. 3 (e).

Using a network analyzer, we are able to measure the electro-optic modulation in the frequency domain. Fig 4 (a) shows the EO modulation amplitude (S$_{21}$) for a micro-ring with a quality factor of 80,000, revealing a 3 dB electrical bandwidth of 2.3 GHz. The inset shows the resonance used in obtaining the frequency response curve. Generally, for resonator based modulators, the modulation roll-off at high frequency can be affected by (1) the resistance-capacitance (RC) time constant determined by electrodes and (2) the cavity photon lifetime (proportional to $Q$). Here, for our device, the capacitance of the lumped electrode is measured to be around 10 fF. Therefore the RC time constant is around 1 ps, which is less likely to be the main limiting factor of the cut-off frequency. The cavity photon lifetime, on the other hand, calculated by $\tau_{ph} = \lambda Q/2\pi c$, ($\lambda$ is the wavelength, $Q$ is the quality factor, $c$ is the light speed in vacuum) is around 66 ps for a Q factor of 80,000. As such, the cavity photon lifetime limited cut-off frequency $f_{3-dB} = 1/2\pi\tau_{ph} = 2.4$ GHz is the main limiting factor of our modulator's frequency response, as supported by our network analyzer measurement.

In addition to operating the AlN microring resonators at telecom C-band wavelengths, we fabricate and measure microring resonators with a diameter of 80 μm working in the visible wavelength range from 766 nm to 780 nm. Efficient light coupling in and out of the chip can still be achieved using grating couplers with adjusted grating period and filling factor. The typical coupling efficiency is 10% per grating coupler for 770 nm light. In order to ensure single mode operation, the visible ring resonators are built on 330 nm thick AlN films instead of 650 nm thickness (used for the 1550 nm light) and the waveguide width is reduced to 400 nm. Coherent input light is launched from a diode laser source (New Focus 6712) and guided to the fiber array which also contains 770 nm single mode polarization maintaining fibers. The output light is detected by a low-noise silicon photoreceiver (New Focus 2107). By varying the coupling gap, we can achieve near critical coupling with measured quality factor of 12,000. For weakly coupled microrings, the fitting reveals a quality factor of 20,000, which corresponds to a propagation loss of 20 dB/cm.

Similar to the approach employed for infrared modulation, we set the input wavelength to a resonance near 770.5 nm, and apply the same binary test sequence at 1 Gb/s (Fig. 3 (a)) as used for infrared modulation on the visible light microrings. The output light is sent to a 12.5 GHz GaAs photodetector (818-BB-45F). The electrical output from the photodiode is electrically amplified and shown in Fig 3 (c) (due to the relatively higher loss and lower efficient detectors at visible bands, the trace shown is an average of multiple measurements). The results show that the modulator also functions effectively at visible wavelengths. In Fig 4 (b) we show the frequency response ($S_{21}$) of the visible light modulation. Because of the lower quality factor (Q=10,000) in the visible, the 3-dB cutoff frequency is limited by the photodiode bandwidth (12.5 GHz) prior to the rolling-off due to the cavity photon lifetime.

The infrared and visible AlN modulators demonstrated in this study will be useful for a myriad of applications involving on-chip broadband phase and amplitude modulations. Using micro-ring resonators to demonstrate the Pockels effect we show that the modulation speed is only limited by the ring cavity linewidth. Higher data rate operation is therefore feasible by employing microrings with lower quality factors. Utilizing state-of-the-art CMOS technology and materials, the modulator's driving voltage can be further reduced by enhancing the effective

RF field inside the AlN waveguides. One strategy is to employ a buried electrode underneath the bottom oxide layer[28] so that a vertical electric field ($E_z$) is present in the AlN waveguides in a parallel plate configuration. Numerical simulations show that the average $E_z$ will receive a five-fold boost compared with the fringing-field induced by planar electrodes. It is also possible to replace the upper PECVD $SiO_2$ with CMOS compatible high-κ oxide, which reduces the effective electric thickness of the cladding materials and hence increases the field strength inside the waveguides according to Fig. 2(c). Furthermore, in case of visible modulators, the top electrode may be replaced by a thin semi-transparent electrode in very close contact with the AlN waveguides given AlN's high breakdown field strength (100 MV/m). These measures would eventually enable AlN modulators with sub-Volt driving voltage and hence be electrically compatible with CMOS circuits.

Recently, power consumption has become an important metric to assess a modulator's performance because modulators contribute a significant part to the total power budget of optical links. It is argued[29, 30] that sub-pJ/bit power consumption needs to be achieved for the implementation of viable on-chip optical networks. In our case, due to the insulating nature of AlN and the cladding materials, DC leakage power is entirely eliminated. Furthermore, because of the tiny electrode capacitance (10 fF), ultra-low energy per bit as small as 10 fJ/bit ($E$/bit= $CV^2/4$) can be estimated. With such a small figure of power consumption, it becomes feasible to build functional optical links working at much higher data rates (such as 40 Gb/s) by employing an array of wavelength domain multiplexed modulators working at lower data rates.

While researchers have been investigating the most suitable Pockels materials for silicon photonics, the true potential of AlN lies in the fact that wafer-scale deposition of high quality films can be conducted in a mature procedure compatible with CMOS manufacturing. The nature of the AlN sputtering process also makes it possible to design three dimensional and multi-layer structures which can enable more flexible and efficient integration. The fully CMOS-compatible AlN modulator is a promising candidate for electro-optic signal processing on a silicon photonics platform. We anticipate AlN-based photonic circuitry could facilitate a plethora of $\chi^{(2)}$ enabled functionalities such as second harmonic generation, parametric downconversion and optical parametric oscillator to be realized on a monolithic platform.

**Acknowledgement**

The authors acknowledge funding support from the NSF CAREER award, the NSF grant through MRSEC DMR 1119826 and the Packard Foundation. W.H.P. Pernice acknowledges support by the DFG grant PE 1832/1-1.The authors thank Dr. Michael Rooks and Michael Power for assistance in device fabrication.

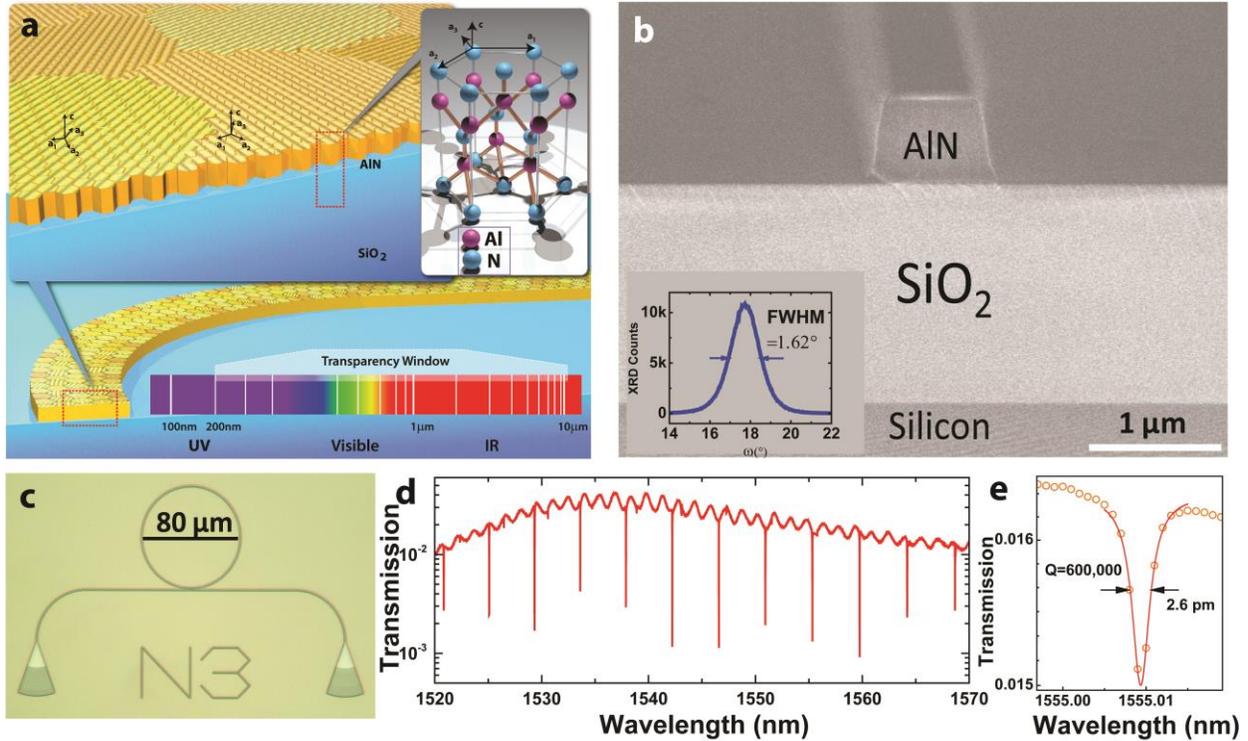

**Figure 1. (a)** The sputtered AlN thin films (in yellow) are polycrystalline, consisting of columnar micrograins with c-axis oriented out-of-plane. The micrograins have randomized in-plane orientations illustrated by the hatched lines on the surface. Inset: atomic arrangement of a hexagonal AlN unit cell. AlN's ultra-wide bandgap (6.2 eV) allows an optical transparency window spanning from UV to infrared. **(b)** A cross sectional SEM image of a patterned AlN waveguide sitting on 2 μm thick thermally grown $SiO_2$ on silicon substrates. The inset shows a typical X-ray rocking curve measurement of the AlN (0002) peak. The full width at half maximum (FWHM) is about 1.62°, indicating excellent c-axis orientation. **(c)** An optical micrograph of the measured AlN ring resonator with a diameter $d$=80 μm. **(d)** Optical transmission spectrum of a microring resonator working in the 1550 nm band with a coupling gap of 300 nm. The free spectral range (FSR) is about 4.3 nm and loaded quality factor is about 100,000. **(e)** An undercoupled ring resonator showing a resonance at 1555.01 nm with a linewidth of 2.6 pm, corresponding to a Q of 600,000.

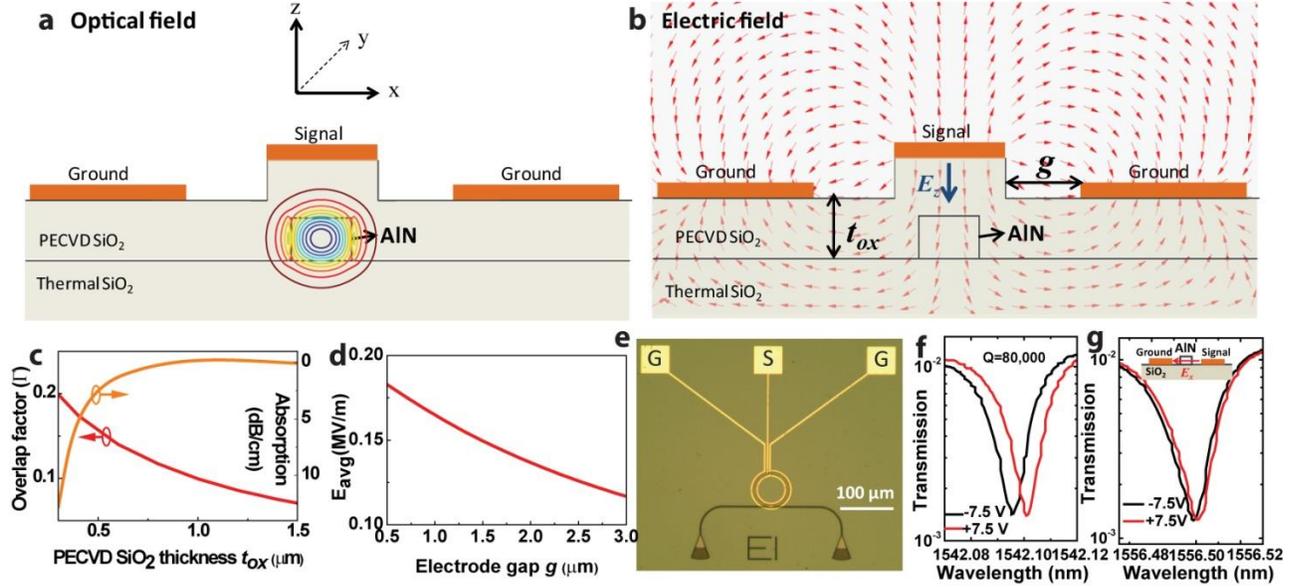

**Figure 2.** (a) The numerically calculated fundamental TE-like mode in AlN waveguide (cross sectional view). Shown are contour plots of electric field, in-plane component. (b) The numerically calculated electric field distribution induced by the Ground-Signal-Ground electrodes. The arrows denote the direction of the applied electric field. The most effective electric field contributing to modulation is the out-of-plane component $E_z$. The electrode gap ($g$) and the deposited oxide thickness ($t_{ox}$) are varied to achieve the largest electro-optic integral ($\Gamma$) inside the waveguides. (c) For a fixed electrode gap $g=2$ μm, we show the numerically calculated electro-optic overlap integral ($\Gamma$) and the optical absorption due to the electrodes versus the PECVD SiO$_2$ thickness ($t_{ox}$). An optimal oxide deposition thickness of 0.8 μm is used in the experiment for the highest electric field inside the waveguides and simultaneously small absorption (< 1 dB/cm) from the electrodes. (d) The average z component of RF field ($E_z$) decreases as the electrode gap ($g$) widens. The simulation was carried out at a fixed deposited SiO$_2$ thickness of 0.8 μm and +1 V bias voltage on the signal electrode. (e) A microscopic image of a microring modulator working at telecom wavelengths. A high frequency GSG probe is used to probe the three contact pads. A pair of grating couplers near the bottom of the image is employed to couple light in and out of the chip. (f) A ring resonance near 1542.10 nm with an extinction ratio of 10 dB and quality factor of 80,000 is tuned 8 pm by applying dc bias from -7.5 V to +7.5V. (g) Measurement of electro-optic tuning of one resonance at 1556.50 nm using EO coefficient $r_{51}$. The inset shows a schematic of the placement of two electrodes by the side of the AlN waveguide which introduce an in-plane electric field inside the waveguides.

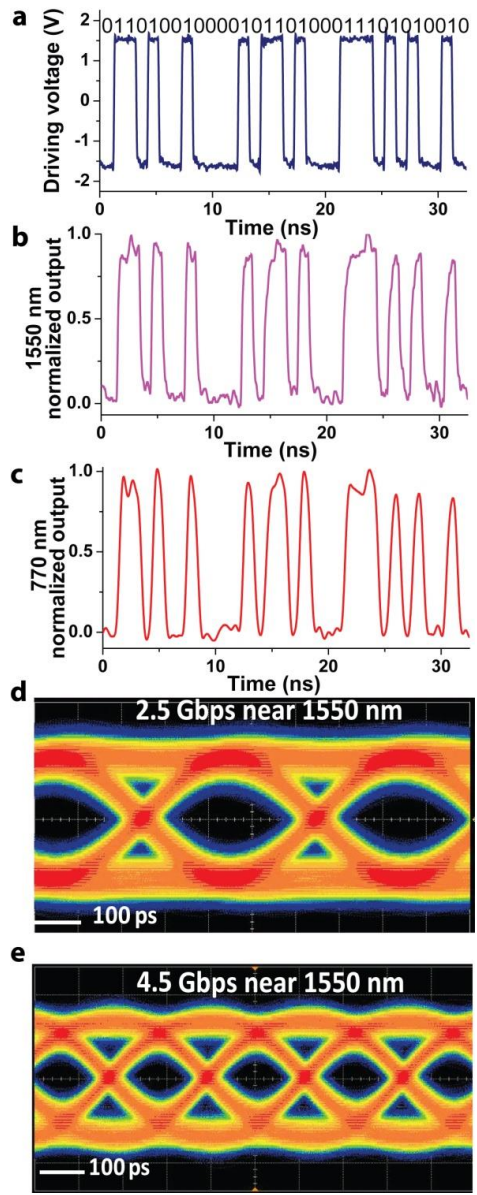

**Figure 3.** **(a)** Input waveform of a 32-bit non-return-to-zero (NRZ) binary sequence ("01101001…0010") at 1 Gb/s with a peak to peak voltage of 6.6 V applied to the test the modulator. **(b)** The normalized output from the telecom band AlN ring modulator for the input test sequence in (a). **(c)** The normalized output (after multiple averages) from the visible light ring modulator for the input test sequence in (a). With pseudo-random binary sequence (PRBS) electrical input, eye diagrams are recorded at 2.5 Gb/s **(d)** and 4.5 Gb/s **(e)** respectively (telecom band). The eye extinction ratio is 3 dB with an applied peak to peak voltage of 4 V.

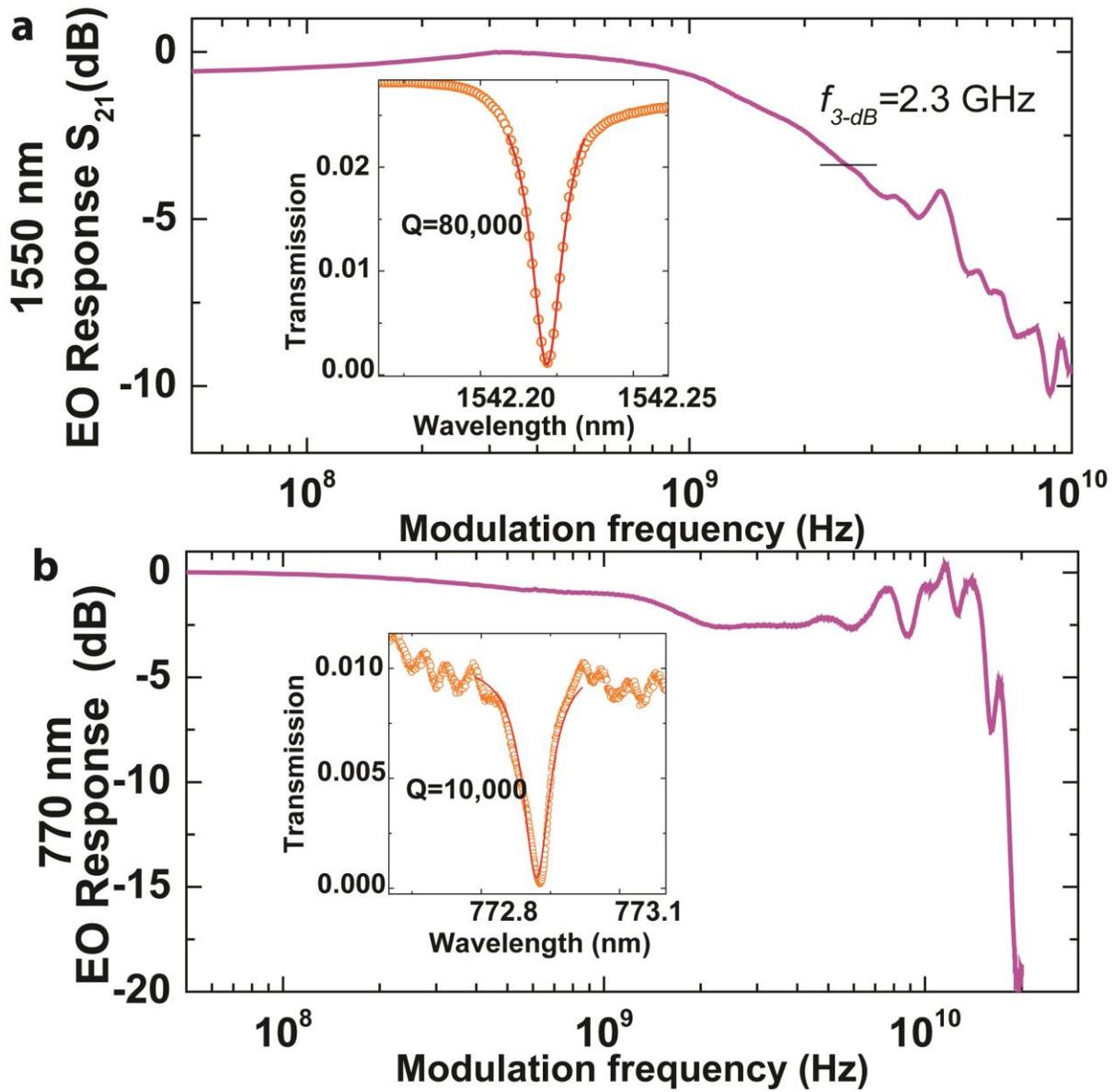

**Figure 4.** **(a)** and **(b)**: The frequency response of the electro-optic modulation at telecom wavelengths and visible wavelengths respectively. Insets show the optical resonances utilized for modulation operations. The 1550 nm modulator has a 3-dB cutoff frequency near 2.3 GHz, which is limited by the cavity photon lifetime. The visible light modulator has a higher cutoff frequency because of the lower Q in the visible. The roll-off around 10 GHz is due to the bandwidth limit of the photodetector.